# Field Label Prediction for Autofill in Web Browsers

Joy Bose
Microsoft IDC
Hyderabad, India
joy.bose@ieee.org

*Abstract*—Automatic form fill is an important productivity related feature present in major web browsers, which predicts the field labels of a web form and automatically fills values in a new form based on the values previously filled for the same field in other forms. This feature increases the convenience and efficiency of users who have to fill similar information in fields in multiple forms. In this paper we describe a machine learning solution for predicting the form field labels, implemented as a web service using Azure ML Studio.

## I. INTRODUCTION

Automatic form fill [1-3] is a feature in web browsers where the fields in a web form are filled automatically upon loading of the form. This works by predicting the field labels and automatically suggesting or filling the previously stored information for those fields, based on the user's historical data stored locally in the browser. This feature is present in all major web browsers and results in productivity enhancements since the user does not have to fill the same form field repeatedly for multiple forms. In order for this feature to work, the field labels for a new form would have to be predicted correctly. This is the problem we are seeking to solve in this paper.

A naïve approach for predicting the form labels is to keep a file with the extracted features and the predicted form labels for each field in each form. However, such an approach would not be scalable, since the web has billions of forms.

Chrome [1], Firefox and other major browsers use a combination of heuristic rules for this purpose. An example of such a rule may be: if the Id or name or label of the field HTML follows a specific regular expression, then the predicted field should be set to a specific value. However, such an approach may not give desired results for some kinds of forms where the Id may be ambiguous or for forms that are dynamic, such as forms behind a login or paywall.

In such a case, a machine learning based approach, based on predicting the fields of a web form based on learning using a labeled dataset, might give better results.

In this paper we describe the design and implementation of a machine learning solution, comprising a web service to predict the field labels for the fields in a given web form. A machine learning model is trained on the server, and predicts the label of the form field in real time given a set of features extracted from the HTML of the web form. Alternatively, the trained model file can also be included as part of the web browser executable or by using a browser extension. However, we focus on the web service approach in this paper since it is easier to update the model when new data is acquired. It must be noted that the actual form data remains in the web browser client since that data is private, only the features related to the form and field details extracted from the HTML are sent to the server and used to predict the field labels.

Fig. 1 shows a sample autofill interface. Fig. 2 shows the steps of training the machine learning model, and fig. 3 shows the architecture of the web service to predict the field labels.

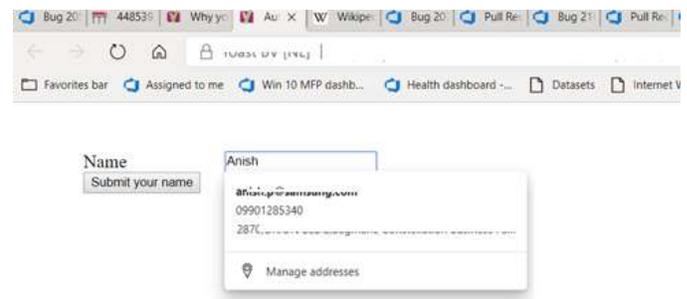

Fig. 1. A sample interface for autofill suggestions for a web form

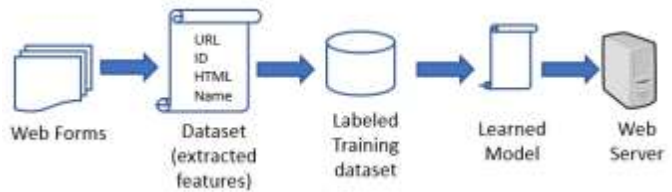

Fig. 2. Steps for training the machine learning model to predict form fields

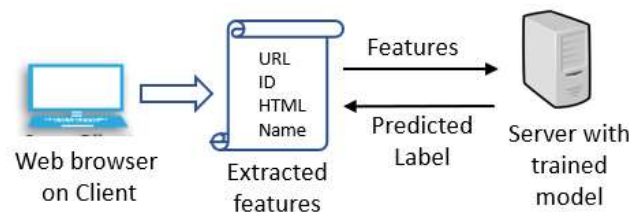

Fig. 3. Architecture of the web service to predict the labels for form fields

## II. RELATED WORK

The autofill feature in browsers [1-3] has been around for a long time, and a number of patents [4-6] also exist in this area.

Liddle [7] built an early prototype for extracting user data for auto form fill. They explored the feasibility of getting the relevant user information to enable autofill, experimenting with different queries. However, this solution was proposed before autofill was widely available in major web browsers, so it concentrates on the feasibility rather than the performance accuracy.

Winckler et. al [8] proposed a solution to autofill that explored getting user information from different sources of user data with varying levels of privacy.

Hartmann [9] explored the feasibility of building an auto form fill solution that is context aware, using a mapping between a context store and the user interface of the form to get more accurate labels.

Wang [10] analyzed web user interface components, using clustering to find semantically similar UIs to enable autofill rather than using fixed labels as used in major web browsers.

However, the above solutions did not focus more on the practical issues encountered in a large scale implementation. They also did not directly use a machine learning solution to classify the labels of the form fields, as is the focus of this paper.

In the following sections, we detail the steps of preparing the dataset and training the machine learning model.

## III. DATASET PREPARATION

In order to train a machine learning model to predict the field labels in a web form, the first step is to generate a dataset. This is done by extracting features from the HTML of multiple web forms and manually providing a label for each of the fields. Common labels used in most web forms are name, address, username, password, age etc.

We used a crowdsourced method with human labelers to generate a dataset of around 4000 values from commonly used web forms.

We use label, name, id and URL as the input features. We performed some basic preprocessing to remove stop words etc. For each of the form fields, we train a separate binary classifier, although one can also experiment with a multi-class classification approach.

For encoding the input dataset as a feature vector, we used a one hot encoding approach, where we got a dictionary of possible values and each of the features was a value in that dictionary. For example, we got all the possible field names from our dataset (such as username, username_01 etc) and built a feature vector. The final feature vector is built by concatenating the individual encodings of the different features.

Fig. 4 shows a sample dataset for an email label predictor.

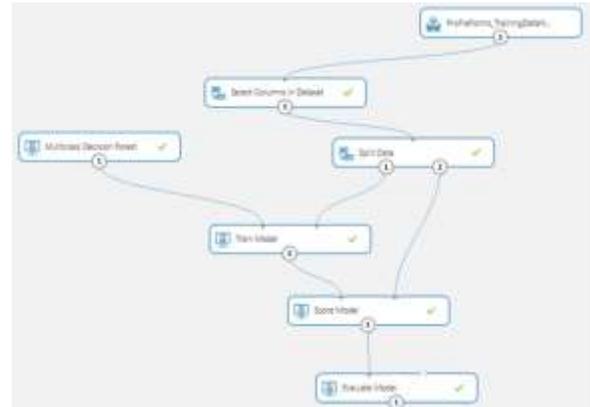

Fig. 4. A sample dataset for predicting whether the label is email. The last column is the predicted label as a binary value

## IV. CLASSIFICATION USING AZURE ML STUDIO

Azure Machine Learning (ML) studio [11] provides a simple and easy to use interface for training and deploying machine learning models. One has to simply drag and drop modules from a menu, which can also include custom code and draw connections between the different modules.

Fig. 5. Interface for training a machine learning model using Azure ML studio

The Azure ML studio gives features for ingesting data in different input forms such as CSV, different types of input preprocessing, experimenting with different types of models along with scoring and evaluation of models. An additional advantage, aside from ease of use, of the studio is that it can easily convert the model to a web service hosted on Azure and provide the APIs to call in order to get the result from the web service. That is why we preferred to use it. However, one may use any other machine learning platform and get the same results with similar models and parameters.

Using the Azure ML studio, we trained a model on our dataset and experimented with different machine learning algorithms and parameters, trying to optimize the accuracy. Once we had obtained a desired level of accuracy for each of the labels (email classifier, state classifier, or multi class classifier etc), we exposed the model as a web service to call from the web browser client using a browser extension.

Fig. 5 gives the screenshot of the interface for a field label predictor using Azure ML studio.

## V. MODEL AND PRELIMINARY RESULTS

We configured a train:test ratio of 70:30 for our solution.

After experimenting with different models including linear regression, support vector machines and decision trees, we found that a multi class decision forest (one of the off-the-shelf ML algorithm libraries available in the Azure ML studio) gave the best results for our dataset.

The parameters we used for the decision forest are as follows: resampling method = bagging, number of decision trees = 16, maximum depth = 100, random splits per node = 128, maximum samples per leaf node = 1.

After tuning the model parameters to improve the accuracy, we obtained an overall precision of around 95% for the email classifier. We obtained comparable results for the multi-class classifier as well.

## VI. CONCLUSION AND FUTURE WORK

In this paper, we discussed a machine learning based solution for autofill feature of web browsers. We trained the model and implemented the system as a web service using the Azure machine learning studio, and obtained good results for our form field label classification.

In future, we intend to train our model on a bigger and more varied dataset. We also plan to experiment with a hybrid approach based on a ensemble of different approaches (lookup table, regular expressions and machine learning). Such a hybrid approach might be expected to work better for most cases than any of the individual approaches.